\documentclass[12pt]{iopart}

\usepackage{amsfonts,amsbsy,amssymb}
\usepackage{mathrsfs,bbm}
\usepackage{graphicx}

\newcommand{\ket}[1]{| #1 \rangle}
\newcommand{\expect}[1]{\langle #1 \rangle}
\newcommand{\bra}[1]{\langle #1 |}
\newcommand{\abs}[1]{| #1 |}

\newcommand{\ketbra}[2]{| #1 \rangle \langle #2|}
\newcommand{\braket}[2]{\langle #1 | #2 \rangle}

\renewcommand{\tr}[1]{\mathrm{tr}[#1]}

\newcommand{\St}{{\hat{S}_t}}
\newcommand{\Sqt}{\hat{S}(\xi_t)}
\newcommand{\Yt}{\hat{R}_\mathrm{y}(\theta_t)}

\newcommand{\partialD}[2]{\frac{\partial #1}{\partial #2}}

\newcommand{\Fx}{\hat{F}_{\mathrm{x}}}
\newcommand{\Fy}{\hat{F}_{\mathrm{y}}}
\newcommand{\Fz}{\hat{F}_{\mathrm{z}}}
\newcommand{\Fk}{\hat{F}_{k}}
\newcommand{\Fpx}{\hat{F}_{+,x}}
\newcommand{\Fmx}{\hat{F}_{-,x}}

\def\sHOmegaZero{\hat{s}^{\phantom{\dagger}}_{0,\omega}}

\def\sHOmegaX{\hat{s}^{\phantom{\dagger}}_{\mathrm{x},\omega}}

\def\sHOmegaY{\hat{s}^{\phantom{\dagger}}_{\mathrm{y},\omega}}

\def\sHOmegaZ{\hat{s}^{\phantom{\dagger}}_{\mathrm{z},\omega}}

\def\aDOmegaX{\hat{a}^\dagger_{\mathrm{x},\omega}}
\def\aHOmegaX{\hat{a}^{\phantom{\dagger}}_{\mathrm{x},\omega}}
\def\aDOmegaY{\hat{a}^\dagger_{\mathrm{y},\omega}}
\def\aHOmegaY{\hat{a}^{\phantom{\dagger}}_{\mathrm{y},\omega}}

\begin{document}

\title[Coherent Feedback in Collective Spin Systems]{Amplified Quantum Dynamics  and Enhanced Parameter Sensitivity via Coherent Feedback  in Collective Atomic Spin Systems}
\author{Bradley A. Chase and JM Geremia}
\ead{\mailto{bchase@unm.edu}, \mailto{jgeremia@unm.edu}}
\address{Quantum Measurement \& Control Group, Department of Physics \& Astronomy, The University of New Mexico, Albuquerque, New Mexico 87131 USA}
\begin{abstract}
We consider the effective dynamics obtained by double-passing a far-detuned laser probe through a large atomic spin system.  The net result of the atom-field interaction is a type of coherent positive feedback that amplifies the values of selected spin observables.   An effective equation of motion for the atomic system is presented, and an approximate 2-parameter model of the dynamics is developed that should provide a viable approach to modeling even the extremely large spin systems, with $N \gg 1$ atoms, encountered under typical laboratory conditions.  

Combining the nonlinear dynamics that result from the positive feedback with continuous observation of the atomic spin offers an improvement in quantum parameter estimation.  We explore the possibility of reaching the Heisenberg uncertainty scaling in atomic magnetometry without the need for any appreciable spin-squeezing by analyzing our system via the quantum Cram\'{e}r-Rao inequality.  Finally, we develop a realistic quantum parameter estimator for atomic magnetometry that is based on a two-parameter family of Gaussian states and investigate the performance of this estimator through numerical simulations.   In doing so, we identify several issues, such as numerical convergence and the reudction of estimator bias, that must be addressed when incorporating our parameter estimation methods into an actual laboratory setting. 

\end{abstract}

\maketitle

\section{Introduction}

Real-time feedback plays an essential role in controlling the behavior of physical systems that must respond to unanticipated events on the same time-scale as their own dynamics \cite{Doyle:1990a,Doyle:1997a,Jacobs:1996a}.  This basic engineering premise shows up everywhere: mechanical actuation based on situational detection provides the basis for devices like the air handling system in a typical optics lab as well as the autopilot on a typical airliner; electrical feedback lies at the heart of error correction in communication systems; and biochemical feedback implements metabolic control and maintains homeostasis in just about every living organism.   Another example, and one which plays an important role in this work, involves the amplification of signals using positive feedback \cite{Wiener:1948a}.  In the case of amplification, feedback must be implemented in real-time because a bona fide amplifying device must operate on signals whose form and value is not  known in advance.

Engineering quantum mechanical systems shares much in common with its classical counterpart in that optimal \cite{Pierce:1998a,James:2004a} and real-time feedback also provides an essential tool for controlling behavior in the face of unpredictable fluctuations.  Examples include stabilizing Heiseneberg uncertainty to achieve deterministic state preparation \cite{vanHandel:2005a,Thomsen:2002a}, implementing optimal measurements for quantum state discrimination \cite{Cook:2007a}, quantum error correction \cite{Ahn:2002a}, and affecting the motion \cite{Fischer:2002a,Durso:2003a} or internal states \cite{Smith:2002a} of single particles or qubits, to name a few.  Quantum feedback does differ from its classical analog, however.   In the classical picture of feedback, one continually measures the system to determine its deviation from the intended behavior by constructing an \textit{error signal} and then manipulates the system to put it back on track by minimizing the error.   Every part of that process is governed by the same type of physical laws, so to speak; the measurements, actuation, external influences, etc., can all be modeled using classical mechanics, thermodynamics, information theory, etc.  The situation is different in quantum mechanics, where the system undergoes substantially different types of dynamics depending on whether it is being observed or not, which is significant because it implies that there are qualitatively different ways to enact feedback.   Quantum feedback control is therefore typically broken down into different forms:
\begin{enumerate}
\item \textit{Measurement-based feedback:} continuous observation of the quantum system is performed by entangling it with an auxiliary meter, such as the electromagnetic field, and measuring the meter continuously in time to gain information about the target system \cite{Belavkin:1983a,vanHandel:2005a}.   A control Hamiltonian is then modulated by changing the values of classical parameters (such as field strengths) in the Hamiltonian as a time-dependent function based on the evolving measurement record \cite{Wiseman:1994a}.  The feedback controller is implemented using classical signal processing techniques to interpret the continuous measurement record and determine the appropriate feedback signal from a classical control law.  Feedback is applied by modulating the system's quantum Hamiltonian as a function of time.

\item \textit{Coherent feedback:} the quantum system of interest is coupled coherently to an auxiliary quantum system according to some Hamiltonian.  The auxiliary system directly plays the role of the feedback controller \cite{Lloyd:2000a,Mabuchi:2008a}.  Its dynamical state changes based upon that of the target system as a result of the coherent interaction that couples them, and the state of the system likewise depends upon that of the controller (either through the same interaction, or some additional coupling).  In this sense, the dynamics of the target system are modified based on its own state via the external coupling.   The coupling Hamiltonian responsible for the feedback is generally not modified as a function of time.
\end{enumerate}
In this paper, we consider a hybrid type of control process that combines aspects from both of the two varieties of feedback just described.  A laser field that is double-passed through an atomic spin system (q.v., Figure \ref{figure:schematic}) \cite{Sherson:2006a,Sarma:2008a,Muschik:2006a,Chase:2009a,Chase:2009b} provides a type of coherent feedback interaction.  On the first pass of the light through the atoms, the polarization state of the field is modified in a manner that depends upon the atomic spin.  Then on the second pass, the interaction is configured such that the light drives the atoms in a way that depends upon the optical polarization.  The net result is a positive-feedback amplification of selected atomic spin observables.   Both of the atom-field interactions are coherent in nature, but the field is also an infinite reservoir that would decohere the atomic sample if otherwise neglected.  Combining the atom-field interaction with continuous measurement of the twice-scattered field, however, helps to limit this decoherence.  Measurement also enables the estimation of atomic spin observables, which in turn can be used to estimate the strength of external influences on the atomic spin.   To demonstrate the utility of this system for quantum parameter estimation, we consider the problem of determining the magnitude of an external magnetic field.  The field estimation procedure is designed in such a way that the positive coherent feedback caused by the optical field amplifies the spin dynamics that result from the magnetic field in a manner that improves the overall sensitivity \cite{Chase:2009a,Chase:2009b}.

The remainder of the paper is organized as follows. Section (\ref{section:physical_system}) develops an equation of motion for the coherent atom-field feedback interaction in the form of a stochastic time-evolution operator for the joint system.  Section (\ref{section:quantum_filter}) develops the so-called \textit{quantum filter} that allows one to estimate atomic observables based on continuous detection of the scattered field. Section (\ref{section:parameter_estimation}) utilizes the filtering equations to develop an estimator for determining the strength of an applied magnetic field based on the induced dynamics of the atomic system in the combined presence of the magnetic and coherent feedback interactions.

\section{The Physical System} \label{section:physical_system}

\begin{figure}
\begin{center}
\includegraphics{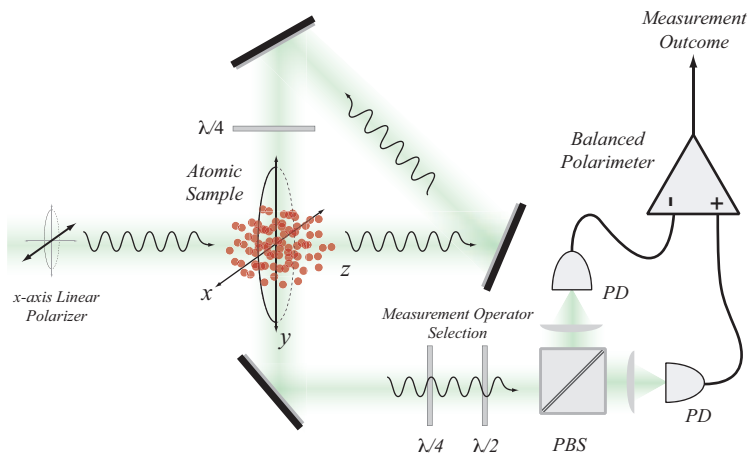}
\end{center}
\vspace{-4mm}
\caption{(color online) The collective spin angular momentum of $N$ atomic spins is coupled to a far-detuned laser field that is double-passed through the atomic sample.  On the first pass through the atoms, the field is linearly polarized and therefore acquires a Faraday rotation proportional to the $z$-component of the atomic spin.  On the second pass, the Faraday rotation is transformed into optical helicity, which the atoms perceive as a fictitious magnetic field.  Symbol defintitions: (PBS) polarizing beam splitter, (PD) photodetector, ($\lambda/2$) half-waveplate, ($\lambda/4$) quarter-waveplate. \label{figure:schematic}}
\end{figure}

Consider a physical system comprised of a cloud of $N$ atoms coupled to the electromagnetic field, as depicted by the schematic in Fig.\ (\ref{figure:schematic}).   The degrees of freedom of interest are provided by the hyperfine angular momentum of the atomic system, which arises from the coupling of the spin and orbital angular momentum of the atom's valence electrons with its nuclear spin
\begin{equation}
	\hat{\mathbf{f}}^{(j)} = \hat{\mathbf{s}}^{(j)} + \hat{\mathbf{l}}^{(j)} + \hat{\mathbf{i}}^{(j)}.
\end{equation}
Here, the superscript $j = 1, \ldots, N$ indicates on which of the $N$ atoms the corresponding operator acts.  A basis for the atomic Hilbert space $\mathscr{H}_\mathrm{A}$ is then provided by the simultaneous eigenkets of the total spin and its $z$-projection 
\begin{eqnarray}
	(\hat{f}^{(j)} )^2 \ket{f_j,m_j} & = & \hbar^2 f_j(f_j+1) \ket{f_j,m_j} \\
	\hat{f}_\mathrm{z}^{(j)}\ket{f_j,m_j} & = & \hbar m_j \ket{f_j,m_j}  
\end{eqnarray}
for each of the $j=1,\ldots, N$ atoms.  In many laboratory situations, however, the dynamic quantum state of the atoms resides in a small sub-space of $\mathscr{H}_\mathrm{A}$ called the \textit{completely symmetric representation}.   Such states are obtained when the total spin quantum number is the same for all of the atoms $f_j=f_{j'} = f$ and the state of the $N$ atoms remains invariant under a permutation of particle labels $\mathcal{P}_{j,j'}[ \hat{\rho} ] = \hat{\rho}$ \cite{Chase:2008a,Stockton:2003a}.  The effective atomic Hilbert space is then spanned by the eigenkets
\begin{equation}
	\hat{F}_\mathrm{z} \ket{F, M}  =  \hbar M \ket{ F,M}, \quad M = -F, \ldots, +F
\end{equation}
of the \textit{collective spin operators}
\begin{equation}
	\hat{F}_i = \sum_{j=1}^N \hat{f}_i^{(j)}.
\end{equation}
Dynamics that are generated by functions of the collective spin operators preserve the total spin quantum number, and for spin-polarized atomic systems such as those considered here, its value corresponds to the maximal angular momentum states $F = N f$.  The dimension of the symmetric atomic Hilbert space is therefore linear in the number of atoms $g_F = 2 N f + 1$, and the atomic state will remain confined to this sub-space provided that the dynamics do not distinguish between the different atoms \cite{Chase:2008a}.  In this paper, only states and dynamical models that preserve symmetric collective states to good approximation are considered \cite{Chase:2008a}.   In practice, realizing such a model in a laboratory setting would require that the electromagnetic fields used to manipulate the atomic sample couple identically to each atom and that processes such as spontaneous emission, which do no preserve the completely symmetric representation, are heavily suppressed \cite{Chase:2008a}. 

\subsection{Atom-Field Coupling}

As also depicted in Fig.\ (\ref{figure:schematic}), the atomic system is illuminated by a relatively intense, far-detuned laser.  The traveling-wave electromagnetic field propagates initially along the atomic $z$-axis and is $x$-polarized, i.e., it is linearly polarized such that its electric field oscillates along the atomic $x$-axis.   After its first interaction with the atomic system, the beam is steered such that it passes through the atomic system a second time, along the atomic $y$-axis \cite{Sherson:2006a,Sarma:2008a,Muschik:2006a}.  Prior to this second interaction, the field is passed through polarization optics that are configured in the following manner:  the $x$-polarized component of the field remains unaffected, while the $y$-polarization component is phase-retarded by a quarter wavelength.  Such a transformation is achieved in the laboratory by placing a $\lambda/4$-plate in the beam path with its principal axis oriented along the $x$-axis.  As a result of this configuration, any rotation of the laser polarization during its first pass through the atoms is converted into optical helicity on the second pass.  The atoms perceive this optical helicity as a fictitious magnetic field, which drives rotations of the collective atomic angular momentum.

Modeling the dynamics of the double-pass atom-field system requires that we develop an equation of motion for the joint state of the atoms and the field, for example by computing the time-evolution operator $\hat{U}_t$ that brings an initial state $\hat{\rho}_0$ to the evolved state $\hat{\rho}_t = \hat{U}_t^\dagger \hat{\rho}_0 \hat{U}_t$ at time $t$.  To do so, we proceed according to the following steps:
\begin{enumerate}

\item a separate interaction Hamiltonian for each of the two passes of the far-detuned laser field through the atomic system is developed;

\item a weak coupling limit is taken to obtain a Markov equation of motion generated by each of the separate interaction Hamiltonians;

\item the two Markov-limit propagators are combined into a single coarse-grained Markov limit that describes the dynamics on time-scales that are slow compared to the time required for the laser field to propagate twice through the atomic system. 
\end{enumerate}
Our intention here is not to present a detailed derivation of the stochastic time-evolution operator and quantum filtering equations that result from the procedure just described, but rather to demonstrate the capabilities of the measurement configuration in Fig. (\ref{figure:schematic}).   For details on the derivations, we refer the interested reader to Ref.\ \cite{Chase:2009b}.

 %%%%%%%%%%%%%%%%%%%%%%%%%%
\subsection{The Hamiltonian for the Atom-Field System}

When the collective spin angular momentum of a multilevel atomic system interacts dispersively with a traveling wave laser field with wavevector $\mathbf{k}$, the atomic spin couples to the two polarization modes of the electromagnetic field transverse to $\mathbf{k}$ according to the atomic-polarizability Hamiltonian 
\begin{eqnarray}
	\hat{H}_\mathrm{AL} & \approx & \sum_{j=1}^N
	\hat{\mathbf{E}} (\mathbf{r}_j, t ) \cdot 
	\left[ \frac{ \hat{\mathbf{d}}^{(j)} \hat{\mathbf{d}}^{(j)\dagger}}{ \hbar \Delta  - i \Gamma / 2}  \right] \cdot 
	\hat{\mathbf{E}} (\mathbf{r}_j, t )  \label{eqn:Hpol1} 
\end{eqnarray}
where $\mathbf{r}_j$ is the position of atom $j$, $\hat{\mathbf{d}}^{(j)}$ is the dipole lowering operator for atom $j$, and $\Delta = \omega_\mathrm{L} - \omega_\mathrm{A}$ is the detuning of the laser from the relevant atomic resonance $\omega_\mathrm{A}$ and $\Gamma$ is the natural linewidth of that resonance.   The atomic polarizability Hamiltonian can be obtained via second-order perturbation theory in which the electronic excited states of the atoms are adiabatically eliminated to obtain an effective Hamiltonian on its ground state \cite{Bouten:2008a,Bouten:2008b}.  

To work with Eq.\ (\ref{eqn:Hpol1}), it is convenient to treat two polarization modes orthogonal to the direction of laser propagation as a Schwinger-Bose field.  When quantized in terms of a plane-wave mode decomposition, one obtains the Stokes operators:
\begin{eqnarray}
 	\sHOmegaZero & = &  +\frac{1}{2} \left(
				\aDOmegaX \aHOmegaX + \aDOmegaY \aHOmegaY 
			\right) \\ 
	 \sHOmegaX & = & +\frac{1}{2} \left(
			\aDOmegaY \aHOmegaY - \aDOmegaX \aHOmegaX	 
	 	\right) \\
	 \sHOmegaY & = & - \frac{1}{2} \left(
	 		\aDOmegaY \aHOmegaX + \aDOmegaX \aHOmegaY
	 	\right) \\
	 \sHOmegaZ & = & + \frac{i}{2} \left( 
	 		\aDOmegaY \aHOmegaX - \aDOmegaX \aHOmegaY
	 	\right), 
 \end{eqnarray} 
expressed in terms of the Schr\"{o}dinger-picture field annihilation operators, $\aHOmegaX$ and $\aHOmegaY$, for the plane-wave modes with frequency $\omega$ and linear polarization along the x- and y-axes, respectively.     But, as is evident from Eq.\ (\ref{eqn:Hpol1}), the interaction Hamiltonian couples the atoms to the electric field as a function of time, suggesting that it would be more convenient to transform the Stokes operators from a plane-wave mode decomposition of the electromagetic field to operators that are labeled by time.  Toward this end, and after moving into an appropriate interaction picture, it is convenient to define the following time-domain Stokes lowering operator 
\begin{equation}
	\hat{s}_t = \frac{1}{2} \int_{-\infty}^{+\infty}  g(\omega) 
		\, \aDOmegaX \aHOmegaY e^{i (\omega-\omega_\mathrm{A}) t} d \omega,
\end{equation}
where $g(\omega)$ is an atom-field coupling parameter that is a function of the field amplitude, detuning, and the atomic transition dipole matrix element \cite{vanHandel:2005a}.  The Stokes operators corresponding to the first-pass and second-pass directions can then be expressed in terms of $\hat{s}_t$ as follows:
\begin{equation}
	\hat{s}_{\mathrm{z},t} =  i 
		\left(  \hat{s}^\dagger_t - \hat{s}^{\phantom{\dagger}}_t \right) \quad\mathrm{and}\quad
	\hat{s}_{\mathrm{y},t}  =  - \left( \hat{s}^{\phantom{\dagger}}_t + \hat{s}^\dagger_t  \right),
\end{equation}
making it possible to view the polarization operators as field quadratures.  

It can be shown that the interaction Hamiltonians for each pass of the probe light through the sample take the form \cite{Bouten:2006a}
\begin{eqnarray} 
		\hat{H}_t^{(1)}  & = & 
			  + i\hbar\lambda_1\Fz \left( \hat{s}_t^{\dagger} - \hat{s}_t^{\phantom{\dagger}} \right)
		\label{eqn:H1} \\
		\hat{H}_t^{(2)}  &= &    -\hbar\lambda_2 \hat{F}_\mathrm{y} \left( \hat{s}_t^{\phantom{\dagger}} + \hat{s}_t^\dagger \right),
		\label{eqn:H2}
\end{eqnarray}
respectively.   The coupling constants $\lambda_1$ and $\lambda_2$ have been introduced to allow one to scale the strength of interaction, and two distinct constants are introduced to account for the possibility that the first and second-pass atom field interaction strengths differ.  In the simplest experimental setting, the laser field will have very nearly the same intensity and detuning on both passes, such that $\lambda_1=\lambda_2$.

Note that in getting from Eq.\ (\ref{eqn:Hpol1}) to Eqs.\ (\ref{eqn:H1}-\ref{eqn:H2}), it was assumed that the traveling-wave nature of the field allows one to average over the atomic position and that rank-two spherical tensor contributions to the dyad $\hat{\mathbf{d}} \hat{\mathbf{d}}^\dagger$ can be neglected \cite{Jessen:2004a,Geremia:2006b}.  In practice, such an approximation depends heavily upon the choice of atomic species, laser detunings, and intensities \cite{Jessen:2004a,Geremia:2006b}.  As spatial field modes other than those corresponding to the direction of the laser propagation have been omitted from the atom-field Hamiltonian, atomic spontaneous emission into non-paraxial modes has implicitly been neglected.  Such an approximation can be justified if the laser is sufficiently far detuned, although in practice, care must be taken to ensure that such conditions are met.\footnote{Neglecting spontaneous emission is justified by the fact that we only consider the case of extremely weak measurements in which the atom-field coupling strength is not large enough to induce appreciable backaction on the atomic system by measuring the forward scattered laser field.}

\subsection{The Stochastic Propagator}

The objective now is to determine the time-evolution of the atom-field system subject to the two Hamiltonians, Eqs.\ (\ref{eqn:H1}-\ref{eqn:H2}), which can be addressed by computing the time evolution operator on the joint atom-field system $\hat{U}_t$.  Toward this end, an equation of motion for the propagator is developed by analyzing the two interaction Hamiltonians separately, and then combining them.  Specifically, this involves calculating the increments $d \hat{U}_t^{(1)}$ and $d \hat{U}_t^{(2)}$ such that the time-evolution operator just after the first pass of the light through the atomic sample is given by
\begin{equation}
	\hat{U}_{t+dt} = \hat{U}_t + d \hat{U}_t^{(1)}
\end{equation}
and the time-evolution operator following the second pass through the atoms is
\begin{equation}
	\hat{U}_{t+2dt} = \hat{U}_{t+dt} + d \hat{U}_t^{(2)}.
\end{equation}
Then, an effective increment $d\hat{U}_t$ is obtained that combines the actions of $d \hat{U}_t^{(1)}$ and $d\hat{U}_t^{(2)}$ into a single update equation for $\hat{U}_t$.

Beginning by analyzing the first pass interaction, the propagator satisfies the equation of motion
\begin{equation}
	\frac{\partial \hat{U}_t}{\partial t}  = - \frac{i \lambda_1}{\hbar}
		 \hat{H}_t^{(1)}  \hat{U}_t =  \lambda_1 \hat{F}_\mathrm{z}  
		\left( \hat{s}_t - \hat{s}_t^\dagger \right)  \hat{U}_t, 
\end{equation}
which can be formally integrated to give
\begin{equation}
	\hat{U}_t = \hat{1} - \lambda_1  \int_0^t d\tau \hat{F}_\mathrm{z} 
		\left( \hat{s}_\tau - \hat{s}_\tau^\dagger \right)  \hat{U}_\tau, 
\end{equation}
from which the increment $\delta \hat{U}^{(1)}_{t} = \hat{U}_{t+\delta} - \hat{U}_t$ can be defined 
\begin{equation} \label{eqn:differential}
	\delta U_t^{(1)} =  \lambda_1   \int_0^{t+\delta} d\tau \hat{F}_\mathrm{z}  
		\left( \hat{s}_\tau - \hat{s}_\tau^\dagger \right) \hat{U}_\tau 
	-    \lambda_1 \int_0^{t} d\tau \hat{F}_\mathrm{z} 
		\left( \hat{s}_\tau - \hat{s}_\tau^\dagger \right) \hat{U}_\tau 	
\end{equation}
and used to obtain a differential equation for $\hat{U}_t$ by taking the limit $\delta \to 0$.

\subsubsection{The van Hove Limit.}

The $\delta \to 0$ limit in Eq.\ (\ref{eqn:differential}) is evaluated by taking the atoms to be weakly coupled to the field, which one would expect to be the case for a finite coupling strength applied over an extremely short period of time $\delta t$.   This so-called ``weak-coupling limit'' is achieved by scaling the perturbation parameter $\lambda \to 0$, however, several technical complications arise in doing so that require one to adopt the methods of quantum stochastic calculus \cite{Accardi:1990a,Gough:2005a,vanHandel:2005a}.   It is, of course, insufficient to take $\lambda \to 0$ simply on its own, as doing so would just turn off the interaction.   The objective of the weak-coupling limit should be viewed in the sense of making a Markov approximation--- it is to determine the effective dynamics of the system on time-scales that are long compared to the correlation time of the field fluctuations. 

In the theory of quantum stochastic calculus, such a limit is obtained by taking $t \to \infty$ at the same time as $\lambda \to 0$ by rescaling time according to $t \mapsto t / \lambda^2$.   This procedure, often called the van Hove limit, has the desired effect of exposing the effective dynamics of the system that result from its average interaction with the rapid fluctuations of the field.  The fact that time should be rescaled as $t/\lambda^2$, and not some other function of $\lambda$, is a byproduct of the correlation function for operators on the field \cite{vanHandel:2005a}.  Indeed, the integrals that arise when taking the weak-coupling limit, such as
\begin{equation}
	 \frac{1}{\lambda} \int_0^\frac{t}{\lambda^2} d\tau \,
		\hat{F}_\mathrm{z} \hat{s}_{\tau/\lambda^2} = 
		\frac{1}{\lambda} \int_0^\frac{t}{\lambda^2} d\tau \int d\omega \, g(\omega)
		\hat{F}_\mathrm{z} 
		\aDOmegaX \aHOmegaY e^{-i (\omega-\omega_\mathrm{A}) \tau/ \lambda^2},
\end{equation}
from Eq.\ (\ref{eqn:differential}) become extremely singular as $\lambda \to 0$.  To get a feel for this singularity, it is useful to consider the commutator
\begin{equation}
	\lim_{\lambda \to 0} \,
		\bigl[ \hat{s}_{t/\lambda^2}, \hat{s}^\dagger_{s/\lambda^2} \bigr] =
	2 \pi | g(\omega_\mathrm{A}) | ^2 \delta( t - s),
\end{equation}
which suggests that the field operators are delta-correlated in time.  Furthermore, the quadrature operators $(\hat{s}_t - \hat{s}_t^\dagger) / i$ commute at different times, which suggests that one should interpret the field operators as white noise, viewed heuristically as arising from the vacuum fluctuations in the probe field.   This analysis can be made mathematically rigorous, and the interested reader is encouraged to consult, for example, Refs.\ \cite{vanHandel:2005a,Bouten:2006a}.

Unfortunately, the presence of white-noise drive terms in Eq.\ (\ref{eqn:differential}) means that the integrals do not converge in the same way that they would were the integrand a smooth function. One must therefore employ the methods of stochastic calculus \cite{Oksendal:1946a} to evaluate the $\delta \to 0$ limit.  By analogy to the procedure adopted in classical stochastic, the extreme singularity of the the quantum white noises $\hat{s}_t$ and $\hat{s}_t^\dagger$, is alleviated by defining the quantum Brownian motion operator
\begin{equation}
	\St = \int_0^t \hat{s}_\tau d\tau
\end{equation}
which enables one to define the the quantum It\^o integral
\begin{equation} \label{eqn:quantum_ito_integral}
	\int_0^t \hat{X}_\tau  d S_\tau = \lim_{\delta \to 0}
		\sum_k \hat{X}_{t_k} \left( S_{t_{k+1}} - S_{t_k} \right)
\end{equation}
where $|t_{k+1} - t_k| = \delta$ and $\hat{X}$ is an arbitrary atomic operator.  The $d\St^\dagger$ and $d \St$ are delta-correlated noise operators derived from the quantum Brownian motion and satisfy the quantum It\^o rules: $d \St d \St^\dagger = dt$ and $d \St^\dagger d \St = d \St^2 = (d \St^\dagger)^2 = 0$.    It is a byproduct of the white-noise singularity that the quantum It\^o integral does not obey the regular chain rule of differential calculus, but rather the It\^o rule
\begin{equation}
	df = \frac{\partial f(X_t,t)}{\partial X_t} dX_t + \frac{\partial f(X_t,t)}{\partial t} dt +
			\frac{1}{2}\frac{\partial^2 f(X_t,t)}{\partial X_t^2} d X_t^2,
\end{equation}
where $f(X_t,t)$ is any well-defined function of $X_t$ and $t$.  This modification of the chain rule is the price one must pay to obtain noise increments that are statistically independent from the state of the system $\mathbbm{E} \int_0^t \hat{X}_\tau  d S_\tau = 0$.   Sometimes however, it is desirable to retain the regular chain rule of calculus,
\begin{equation}
	df = \frac{\partial f(X_t,t)}{\partial X_t} d X_t + \frac{\partial f(X_t,t)}{\partial t} dt,
\end{equation}
such as will be the case when applying the methods of differential geometry to stochastic systems [as will be considered in Sec.\ (\ref{section:projection_filter})].  The regular chain rule of differential calculus can be recovered by defining the stochastic integral as
\begin{equation} \label{eqn:quantum_stratonovich_integral}
	\int_0^t \hat{X}_\tau \circ d S_\tau = \lim_{\delta \to 0}
		\sum_k \frac{1}{2} \left( \hat{X}_{t_{k+1}} - \hat{X}_{t_k} \right) \left( S_{t_{k+1}} - S_{t_k} \right),
\end{equation}
which is called a Stratonovich integral.  However, $\mathbbm{E} \int_0^t \hat{X}_\tau \circ d S_\tau \neq 0$, making it generally less desirable for analysis than its It\^{o} counterpart, Eq.\ (\ref{eqn:quantum_ito_integral}).

With a well-defined quantum stochastic calculus given by Eqs.\ (\ref{eqn:quantum_ito_integral}) and (\ref{eqn:quantum_stratonovich_integral}), it is possible to obtain the first- and second-pass atom-field interactions.  The result of the derivation is the following quantum stochastic differential equations (QSDE) \cite{Chase:2009b}
\begin{eqnarray}
	d\hat{U}_t^{(1)} &=& \left\{\sqrt{m_1}\Fz (dS_t^\dagger - dS_t) 
			- \frac{1}{2}m_1\Fz^2dt 
			- \frac{i}{\hbar} \hat{H}_t dt \right\}U_t^{(1)} \label{eqn:dU1}\\
	d\hat{U}_t^{(2)} &=& \left\{i\sqrt{m_2}\Fy(dS_t + dS^{\dagger}_t) 
			- \frac{1}{2}m_2 \Fy^2dt
			- \frac{i}{\hbar} \hat{H}_t dt \right\}U_t^{(2)} \label{eqn:dU2}
\end{eqnarray}
where $m_i = 2 \pi | \lambda_i g(\omega_\mathrm{A})|^2$.  Here $\hat{H}_t$ is an arbitrary Hamiltonian applied to the atoms that is not involved with their interaction with the probe field.

\subsubsection{Combined Weak-Coupling Propagator.}

To obtain a single weak-coupling limit for the double-pass interaction, the separate equations of motion for the two propagators must be combined into a single weak-couping limit.  To obtain such a coarse-grained propagator, the two single-pass evolutions  $d\hat{U}_t^{(1)} = \hat{a}_t U_t^{(1)}$ and $d\hat{U}_t^{(2)} = \hat{b}_t U_t^{(1)}$ can be concatenated via the expanded differential
\begin{equation}
	d \hat{U}_{t+2 \delta t}  =   \hat{U}_t + \left( \hat{a} + \hat{b} + \hat{b} \hat{a} \right) \hat{U}_t
\end{equation}
such that the combined propagator satisfies $d \hat{U}_t = \hat{U}_{t+2\delta t} - \hat{U}_t$ in the limit where $2\delta t$ is taken to be infinitesimal.\footnote{Care must be taken to account for rescaling time from $2\delta t \rightarrow dt$, as described in Ref.\ \cite{Chase:2009b}}.  After evaluating the combined evolution for the propagators in Eqs.\ (\ref{eqn:dU1}) and (\ref{eqn:dU2}) in light of the quantum It\^{o} rules, the single weak-coupling limit propagator is found to be \cite{Chase:2009b}
\begin{eqnarray} \label{eqn:double_pass_prop}
	d\hat{U}_t  &=&  \left[i\sqrt{KM}\Fy\Fz dt 
		- \frac{1}{2}M\Fz^2 dt - \frac{1}{2}K\Fy^2dt  - \frac{i}{\hbar} \hat{H} dt \right. \\
		 && \left. 
			+ \sqrt{M}\Fz(d\St^{\dagger} - d\St)
			+ i \sqrt{K}\Fy(d\St^{\dagger} + d\St) \right]\hat{U}_t \nonumber
\end{eqnarray}
where $M = m_1 / 2$ and $K= m_2/2$.  Finally, this result can be adapted to the standard notation of quantum stochastic calculus by rewriting the double-pass propagator in the general form 
\begin{equation} \label{eqn:app:propagator}
    d\hat{U}_t = \left[ \hat{L} d\St^{\dag} - \hat{L}^{\dag} d\St
         - \frac{1}{2}\hat{L}^{\dag}\hat{L} dt -i\hat{H}_c dt\right]\hat{U}_t ,
\end{equation}  
with the assignments $\hat{L} = \sqrt{M}\Fz + i\sqrt{K}\Fy$ and $\hat{H}_c = \hat{H} - \sqrt{KM}(\Fz\Fy + \Fy\Fz)/2 $ \cite{Chase:2009b,Sarma:2008a}.

The stochastic propagator allows one to compute the time-evolution of any atomic or field observable.  For example, the Heisenberg-picture operator for any atomic observable $\hat{X}$ (time-independent in the Schr\"{o}dinger picture) is given by the quantum flow $j_t(\hat{X}) = \hat{U}_t^{\dag}(\hat{X} \otimes I)\hat{U}_t$ whose time evolution satisfies 
\begin{equation}
    dj_t(\hat{X}) =  j_t([\hat{X},\hat{L}])d\St^{\dag} 
                      + j_t([\hat{L}^{\dag},\hat{X}])d\St
                      + j_t(\mathcal{L}(\hat{X}))dt \label{eqn:app:dynamics}
\end{equation}            
where $\mathcal{L}$ denotes the familiar Lindblad generator for the operator $\hat{X}$,
\begin{equation}
    \mathcal{L}(\hat{X}) = i[\hat{H},\hat{X}] + \hat{L}^{\dag}\hat{X}\hat{L}
                         - \frac{1}{2}(\hat{L}^{\dag}\hat{L}\hat{X}
                                      +\hat{X}\hat{L}^{\dag}\hat{L}) .
\end{equation}

%%%%%%%%%%%%%%%%%%%%%%%%%%%%%%%%%%%%%%%%%%%%%%%%%%%%%%%%%%%%%%%%%%
\section{Continuous Observation and Quantum Filtering} \label{section:quantum_filter}

In addition to the stochastic propagator $\hat{U}_t$, it is also useful to consider the effect of performing measurements on the twice forward-scattered laser field, as illustrated in Fig.\ (\ref{figure:schematic}).    The task is then to construct a best-estimate of atomic observables conditioned on the measurement data, which can be accomplished using the techniques of quantum filtering theory \cite{vanHandel:2005a,Bouten:2006a}.  Of course, the form of any such filter depends on how the forward scattered probe field is detected.   From the form of the first-pass interaction Hamiltonian $\hat{H}^{(1)}_t$, it is evident that the $z$-component of the atomic spin couples to dynamics generated by the field operator $\hat{s}_{\mathrm{z},t} = i ( \hat{s}^\dagger_t - \hat{s}^{\phantom{\dagger}}_t)$.  The effect of such a coupling would then be observed by measuring the orthogonal quadrature, indicating that information can be gained about the $z$-component of the atomic spin by measuring the polarization observable $\hat{z}_t  =  \hat{s}_{\mathrm{y,t}}$ \cite{Bouten:2007a,Chase:2009b}.    The observation process considered here is therefore described by
\begin{equation}
	Z_t = \hat{U}_t^{\dag}(\St^{\dag} + \St)\hat{U}_t .
\end{equation}  
whose time evolution satisfies the quantum It\^o equation
\begin{equation} 
    \label{eqn:app:observations}
    dZ_t = j_t(\hat{L} + \hat{L}^{\dag})dt 
               + d\St^{\dag} + d\St .
\end{equation}
In the language of classical control theory, Eqs.\ (\ref{eqn:app:dynamics}) and (\ref{eqn:app:observations}) provide the system-observation pair.  The observations process carries information about the state of the system, $j_t(\hat{L} + \hat{L}^{\dag})$, albeit corrupted by the quantum white noise term $d\St^{\dag} + d\St$.  

\subsection{The Quantum Filtering Equation}

Quantum filtering theory is tasked with picking out the relevant information from the observations and combining it with the model of the dynamics in order to estimate system observables.  Mathematically, the filter is given by the conditional expectation 
\begin{equation} \label{eqn:FilterCE}
    \pi_t[\hat{X}] = \mathbbm{E}[j_t(\hat{X}) | Z_{(0,t)}],
\end{equation}
where $Z_{(0,t)}$ is the measurement record up to time $t$.  It can be shown that the conditional expectation in Eq.\ (\ref{eqn:FilterCE}) is adapted to the measurement process and that it is possible to express the evolution of the quantum filter as an It\^o equation
\begin{equation}
    d\pi_t[\hat{X}] = a_t(\hat{X})dt + b_t(\hat{X}) dZ_t ,
\end{equation}
where $a_t(\hat{X})$ and $b_t(\hat{X})$ are functions determined by enforcing $\pi_t$ to satisfy the properties of a conditional expectation, namely
\begin{equation}  \label{eqn:cet}
    \mathbbm{E}[\hat{Y}\mathbbm{E}[\hat{X}|Z_{(0,t)}]]
    = \mathbbm{E}[\hat{Y}\hat{X}]
\end{equation}
for any $\hat{Y}$ that is a function of the measurement process.   Doing so yields the \emph{quantum filtering equation} for the double-pass atomic system \cite{Chase:2009b}
\begin{eqnarray} \label{eqn:app:quantum_filter_eq}
    d\pi_t[\hat{X}]  & = & \pi_t[\mathcal{L}[\hat{X}]]dt  + \left(\pi_t[\hat{L}^{\dag}\hat{X} + \hat{X}\hat{L}]
    - \pi_t[\hat{L}^{\dag} + \hat{L}]\pi_t[\hat{X}]\right) \\
 & &   \quad\quad \quad \quad \times \left(dZ_t - \pi_t[\hat{L} + \hat{L}^{\dag}] dt\right) . \nonumber
\end{eqnarray}
The quantum filter provides the best least-squares estimate of the expectation value of any atomic operator $\hat{X}$ given the measurement data.  This suggests that it should be possible to find a density operator  $\hat{\rho}_t$ such that $\pi_t[\hat{X}] = \tr{\hat{X} \hat{\rho}_t}$ for any observable $\hat{X}$.   Such an operator can indeed be found \cite{Gardiner:1996a,vanHandel:2005a}, and it satisfies the \textit{adjoint form} of the quantum filter
\begin{eqnarray} 
    d\hat{\rho}_t & = & - \frac{i}{\hbar}[ \hat{H}_t,\hat{\rho}_t]dt + i\sqrt{KM}[\Fy,\{\Fz,\hat{\rho}_t\}]dt  
    \label{eqn:adjoint_quantum_filter} \\
	&&    + M \mathcal{D}[\Fz]\hat{\rho}_t dt 
             + K\mathcal{D}[\Fy]\hat{\rho}_t dt 
             + \left(\sqrt{M}\mathcal{M}[\Fz]\hat{\rho}_t 
             + i\sqrt{K}[\Fy,\hat{\rho}_t]\right)dW_t \nonumber
\end{eqnarray}
where the innovations process 
\begin{equation}
	dW_t = dZ_t - 2\sqrt{M}\tr{\Fz\hat{\rho}_t}dt
\end{equation}
is a Wiener process, i.e. $\mathbbm{E}[dW_t] = 0, dW_t^2 = dt$, and the superoperators are defined as
\begin{eqnarray}
    \mathcal{D}[\Fk]\hat{\rho}_t &=& \Fk\hat{\rho}_t\Fk^{\dag} - \frac{1}{2}\Fk^{\dag}\Fk\hat{\rho}_t - \frac{1}{2}\hat{\rho}_t\Fk^{\dag}\Fk\\
    \mathcal{M}[\Fz]\hat{\rho}_t &= &\Fz\hat{\rho}_t + \hat{\rho}_t\Fz - 2\tr{\Fz\hat{\rho}_t}\hat{\rho}_t\\
    \{\Fz,\hat{\rho}_t\} &=& \Fz\hat{\rho}_t + \hat{\rho}_t\Fz .
\end{eqnarray}
In situations where the conditional density operator remains pure under the evolution of the adjoint filter, as is the case for evolution under Eq.\ (\ref{eqn:adjoint_quantum_filter}) provided that the initial state is pure, $\hat{\rho}_t$ can be decomposed as $\hat{\rho}_t = \ket{\psi_t} \bra{\psi_t}$ and shown to satisfy the stochastic Schr\"{odinger} equation (SSE) \cite{Chase:2009a}
\begin{eqnarray}
	d\ket{\psi_t} & = & \mathcal{N}[\ket{\psi}_t] \nonumber \\
	& = & \left(
		 i\sqrt{KM}\Fy(\Fz + \expect{\Fz}_t)
		  -\frac{M}{2}(\Fz-\expect{\Fz}_t)^2  
	 	- \frac{K\Fy^2}{2} - \frac{i \hat{H}_t}{\hbar}
						\right)\ket{\psi_t} dt \nonumber \\
	&&
			  + \left( \sqrt{M}(\Fz - \expect{\Fz}_t) + i\sqrt{K}\Fy\right)\ket{\psi_t} dW_t . 
			   \label{eqn:double_pass_SSE}
\end{eqnarray}
where the superoperator $\mathcal{N}$ is defined as the generator the dynamics.

\subsubsection{Amplification of $y$-Axis Rotations.}

The first term in Eq.\ (\ref{eqn:double_pass_SSE}) suggests the appearance of a Hamiltonian-like term that drives rotations about the $y$-axis in a manner that depends upon the current $z$-component of the atomic spin.  Now consider a system Hamiltonian that involves a $y$-axis rotation
\begin{equation} \label{eqn:y_hamiltonian}
	\hat{H}_t = - \hbar \omega_t \Fy
\end{equation}
where $\omega_t$ is a time-dependent rotation rate about the atomic $y$-axis.  The filtering equation then becomes
\begin{eqnarray}
	d\ket{\psi_t}  & = & \left( i \omega_t \Fy +
		 i\sqrt{KM}\Fy(\Fz + \expect{\Fz}_t)
		  -\frac{M}{2}(\Fz-\expect{\Fz}_t)^2  
	 	- \frac{K\Fy^2}{2} \right)\ket{\psi_t} dt \nonumber \\
	&& \quad	  + \left( \sqrt{M}(\Fz - \expect{\Fz}_t) + i\sqrt{K}\Fy\right)\ket{\psi_t} dW_t . 
			   \label{eqn:double_pass_SSE_2}
\end{eqnarray}
 Since the sign is the same for both the drive and feedback terms (first and second terms in the filtering equation), the resulting feedback is positive and amplifies the effect of the Hamiltonian Eq.\ (\ref{eqn:y_hamiltonian}).    The generator of the feedback dynamics is not, however, Hermitian because the process is not reversible.  The feedback on $\Fy$ is dependent upon $\Fz$ via the interaction of the system with a reservoir.  As in the theory of cascaded quantum systems \cite{Gardiner:2004a}, both the first and second-pass interactions are one-way due to the immensity of the bath. 

\begin{figure}
\begin{center}
\includegraphics{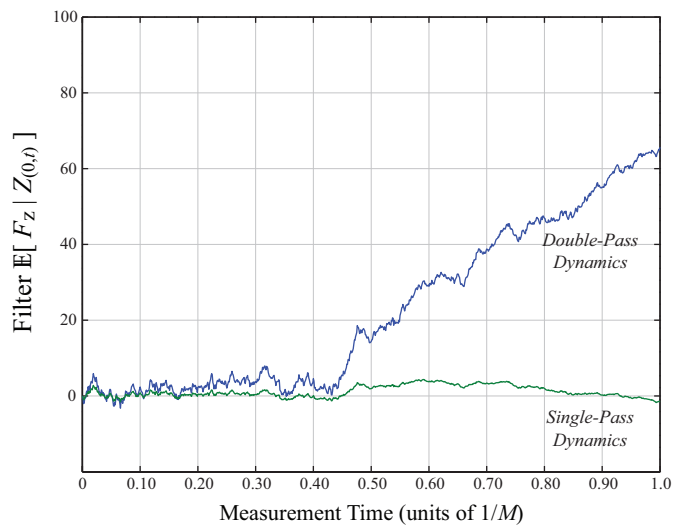}
\end{center}
\vspace{-4mm}
\caption{(color online) Comparison between the atomic dynamics with and without positive feedback amplification enabled.  The measurement trajectories depict the filter $\pi_t(\Fz) = \mathbbm{E}[ \Fz | Z_{(0,t)}] = \bra{\psi_t} \Fz \ket{\psi_t}$ evolved under Eq.\ (\ref{eqn:double_pass_SSE_2}) with $F=100$, $\omega_t = 2.0$ and $M=K=1.7$ (in units of $1/t_\mathrm{final}$).  The double-pass filter shows an amplification of the single-pass dynamics, which were obtained by setting $K=0$ for the sake of comparison.   \label{figure:nonlinear_dynamics}}
\end{figure}

Figure (\ref{figure:nonlinear_dynamics}) illustrates the amplification process just described by comparing the dynamics for a spin system with $F=100$, $\omega_t = 2.0$ and $M=K=1.7$ (with frequencies in units of $1/t_\mathrm{final}$) to one in which the second-pass interaction strength is set to zero $K=0$.   The filtered value $\pi_t(\Fz)$ is plotted for both cases, and the nonlinear effects of the positive feedback are evident from the trajectory.  In order to obtain a meaningful comparison between the evolutions with and without feedback amplification enabled, the filter Eq.\ (\ref{eqn:double_pass_SSE_2}) was propagated using the same noise realization for the innovations process $dW_t$, which is possible in a computer simulation by using the same random seed for both cases.  Of course, the amplification process will also affect the ``quantum noise'' in the atomic system, due to the Heisenberg uncertainty of the initial spin coherent state.  The benefit of amplifying a signal Eq.\ (\ref{eqn:y_hamiltonian}) relative to also amplifying the coherent state projection noise is analyzed in the remaining sections.

\subsection{Reduced Dimensional Filters} \label{section:projection_filter}

While the quantum filter Eq.\ (\ref{eqn:double_pass_SSE}) provides a complete description of the conditional dynamics of the double-pass atomic system subject both to the coherent feedback mediated by field and to continuous observation of the field, it is somewhat impractical to utilize in practice.  Typical experiments involve atomic spin systems with $N \gg 1$ atoms.  Despite that the effective atomic Hilbert space, the completely symmetric representation described in Sec.\ (\ref{section:physical_system}), grows only linearly with $N$, the corresponding operators and kets in Eq.\ (\ref{eqn:double_pass_SSE}) are still too large to be analyzed numerically on a typical high-end computer unless $F$ is on the order of a few hundred or less. 

Preferably, it would be possible to parameterize the atomic Hibert space even further, ideally in a way where the number of parameters does not increase with the number of atoms in the atomic sample.  Such a parameterization is provided by the set of Gaussian states \cite{Duan:2001a,Duan:2002a,Geremia:2003a} that arise when it is possible to truncate the Holstein-Primakoff expansion \cite{Holstein:1940a} of the atomic spin state $\ket{\psi_t}$ to first order (as described below).    Such states arise naturally when an atomic system evolves from an initial spin coherent state, such as the spin-polarized states obtained by optical pumping \cite{Happer:1972a}.  Then, an effective filtering equation can be derived for the small number of parameters that characterize the state.

 \subsubsection{The Family of Gaussian States and Restricted Dynamics.}
 
Such a reduced-dimensional filter can be obtained using the two-parameter family of Gaussian states
\begin{equation} \label{eqn:gaussian_state_parameterization}
        \ket{\theta_t,\xi_t} = \Yt\Sqt\ket{F,+F}_\mathrm{x}, 
\end{equation}
where $\ket{F,+F}_\mathrm{x}$ is the spin coherent state pointing along $+x$, $\Sqt$ is a spin squeezing operator \cite{Kitagawa:1993a} 
\begin{equation}
	\Sqt = \exp\left( -2i\xi_t(\Fz\Fy + \Fy\Fz) \right) 
\end{equation}
with squeezing parameter $\xi_t$ and $\Yt$ is a rotation about the $y$-axis 
\begin{equation}
	\Yt = \exp \left( -i\theta_t\Fy \right)
\end{equation}
by angle $\theta_t$.  This parameterization is, of course, limited; it can only capture dynamics that involve rotations about the atomic $y$-axis and the squeezing that is generated along the $z$-axis that results from measuring $\Fz$.  The rotation via $\Yt$ then accounts for both the random evolution due to the measurement and the feedback term that drives $\Fy$.  

\subsubsection{First-Order Holstein-Primakov States.}

For large $F$, spin-polarized states are extremely well described by the Holstein-Primakoff approximation to lowest order \cite{Holstein:1940a}
\begin{eqnarray} \label{eqn:HP_firstorder}
		\Fpx &\approx &\sqrt{2F} \hat{a}\\
		\Fmx &\approx &\sqrt{2F}\hat{a}^{\dag}\\
		\Fx &\approx& F ,
\end{eqnarray}
where $\hat{F}_{\pm,x} = \Fy \pm i\Fz$, and $\hat{a}^\dagger,\hat{a}$ are bosonic creation and annihilation operators.  The state can then be written as $\ket{F,+F}_\mathrm{x} = \ket{0}$, which is the vacuum in the Holstein-Primakoff representation.  When working with this parameterization, one often encounters inner products of the form
\begin{equation}
    {{}_\mathrm{x}\bra{F,+F}} \Sqt^{\dag}g(\Fx,\Fy,\Fz)\Yt^{\dag} f(\Fx,\Fy,\Fz)
            \Yt\Sqt\ket{F,+F}_\mathrm{x}, \nonumber
\end{equation}
where $g$ and $f$ are polynomial functions of their arguments.  Since $\Yt$ is a rotation, it is possible to evaluate
\begin{equation}
    \Yt^{\dag}f(\Fx,\Fy,\Fz) \Yt 
       = f( \Fx', \Fy', \Fz'), \nonumber
\end{equation}
exactly using 
\begin{eqnarray}
    \Fx' & = & \Yt^{\dag}\Fx\Yt = \Fx \cos{\theta_t} + \Fz\sin{\theta_t} \nonumber \\
    \Fy' & = & \Yt^{\dag}\Fy\Yt = \Fy\\
    \Fz' & = & \Yt^{\dag}\Fz\Yt = \Fz \cos{\theta_t} - \Fx\sin{\theta_t} . \nonumber
\end{eqnarray}
Thus it only remains to evaluate matrix elements of the form
\begin{equation} \label{eqn:hp_expectations}
    {{}_\mathrm{x}\bra{F,+F}}
    \Sqt^{\dag}g(\Fx,\Fy,\Fz)
            f(\Fx',\Fy',\Fz')   \Sqt \ket{F,+F}_\mathrm{x}
\end{equation}
which is possible for small $\xi_t$, i.e., for ``squeezed vacuum'' in the preferred basis.  Under this approximation
\begin{eqnarray}
    \Sqt^{\dag}\Fx\Sqt &=& F\\
    \Sqt^{\dag}\Fy\Sqt &= &\frac{\sqrt{2F}}{2}e^{+4F\xi_t}( \hat{a} + \hat{a}^\dagger)\\
    \Sqt^{\dag}\Fz\Sqt &= &\frac{\sqrt{2F}}{2i }e^{-4F\xi_t}( \hat{a} - \hat{a}^\dagger)  .
\end{eqnarray}
 
\subsubsection{Quantum Projection Filter.}

With the Gaussian family of states defined, an approximate quantum filtering equation confined to the family can be derived.  Our approach utilizes the technique of \emph{projection filtering} \cite{vanHandel:2005b,Mabuchi:2008a}.  Abstractly, the quantum projection filter operates by orthogonally projecting Eq.\ (\ref{eqn:double_pass_SSE}) onto the sub-manifold of states defined by Eq.\ (\ref{eqn:gaussian_state_parameterization}) at each point in time.  In doing so, the objective is to obtain an equation of motion 
\begin{equation} \label{eqn:projlhs}
    d\ket{\xi_t,\theta_t} = v_{\xi_t}d\xi_t + v_{\theta_t}d{\theta_t} .
\end{equation}
in terms of the parameters of the family, namely $\theta_t$ and $\xi_t$, rather than an equation of motion for the full ket $\ket{\psi_t}$.

Given a family of states characterized by the set of parameters $x_1,x_2,\ldots,x_n$ and with states in the family $\ket{x_1,x_2,\ldots,x_n}$, the \textit{tangent space} for that family is spanned by the vectors \cite{vanHandel:2005b}
\begin{equation}
    v_i = \partialD{\ket{x_1,x_2,\ldots, x_n}}{x_i},
\end{equation}
which describe how states within the family $\ket{x_1,x_2,\ldots,x_n}$ are modified under differential changes in the values of the parameters.  Of course the action of the generator $\mathcal{N}[\ket{x_1,x_2,\ldots,x_n}]$ such as that defined by Eq.\ (\ref{eqn:double_pass_SSE}) does not necessarily preserve states in the family.  But, projecting $\mathcal{N}$ onto the tangent space
\begin{equation} 
    T = \sum_i \frac{\langle v_i,
            \mathcal{N}[\ket{x_1,x_2,\ldots,x_n}] \rangle}
                {\langle v_i, v_i \rangle}v_i, \label{eqn:general_projection}
\end{equation}
where the inner product is given by the standard Hilbert space inner product, yields the filter that is the closest approximation to $\mathcal{N}$ that preserves states within the parameterized family.  For the family of Gaussian states defined by, the tangent vectors are found to be \cite{Chase:2009b}
\begin{eqnarray}
    v_{\theta_t} & = & \partialD{\ket{\theta_t,\xi_t}}{\theta_t} =   -i\Fy \Yt\Sqt\ket{F,+F}_\mathrm{x}\\
    v_{\xi_t} & = & \partialD{\ket{\theta_t,\xi_t}}{\xi_t} =
        \Yt\Sqt
        (-2i(\Fz\Fy + \Fy\Fz))\ket{F,+F}_\mathrm{x} .
\end{eqnarray}
with the inner products required for normalization of unit vectors given by \cite{Chase:2009b}
\begin{eqnarray}
     \expect{v_{\theta_t},v_{\theta_t}} &=& \frac{Fe^{8F\xi_t}}{2} \nonumber \\
     \expect{v_{\xi_t},v_{\xi_t}} &=& 8F^2 \label{eqn:tangent_inner_products} \\
     \expect{v_{\xi_t},v_{\theta_t}} &= &0 \nonumber .
\end{eqnarray}
The last line of Eq.\ (\ref{eqn:tangent_inner_products}) indicates that the tangent vectors are orthogonal as required.

\subsubsection{Orthogonal Projection of Double-pass Filter.}

To construct the projection filter onto Gaussian states for the double-pass quantum system in Fig.\ (\ref{figure:schematic}), all that remains is to implement the projections.  Unfortunately, this cannot be accomplished using Eq.\ (\ref{eqn:double_pass_SSE}) directly as the quantum filter in this form is an It\^o equation, whose chain rule is incompatible with the conventional methods of differential geometry \cite{vanHandel:2005a}.  Fortunately, Stratonovich stochastic integrals obey the standard chain rule and are thus amenable to projection filtering methods.  Transforming Eq.\ (\ref{eqn:double_pass_SSE}) to Stratonovich form can be performed using the approach described in, for example, Ref.\ \cite{Chase:2009b} and gives
 \begin{eqnarray} 
	    d\ket{\psi}_t &=& \left[ -i \omega_t \Fy 
	        -M\left[ (\Fz - \expect{\Fz}_t)^2 - \expect{\Delta\Fz^2}_t\right]
		        -\frac{\sqrt{KM}}{2}\Fx \right. \nonumber \\
		     && \left.
		        +2i\sqrt{KM}\expect{\Fz}_t\Fy
		        + i\sqrt{KM} \expect{\Fz\Fy}_t
		        \right]\ket{\psi}_t dt  
		        \label{eqn:double_pass_SSE_stratonovich} \\
		 &&+ \left( \sqrt{M}(\Fz - \expect{\Fz}_t) + i\sqrt{K}\Fy\right)\ket{\psi}_t \circ dW_t
		 	\nonumber ,
\end{eqnarray}
where $\expect{\Delta\Fz^2}_t = \expect{\Fz^2} - \expect{\Fz}^2$.  The projection filter is then obtained by comparing Eqs.\ (\ref{eqn:projlhs}) and (\ref{eqn:general_projection}) using Eq.\ (\ref{eqn:double_pass_SSE_stratonovich}) and the general forms for $d\xi_t$ and  $d\theta_t$
\begin{eqnarray}
    d\theta_t &= &\frac{2e^{-8F\xi_t}}{F} \expect{v_{\theta_t}, d\ket{\psi_t}[\xi_t,\theta_t]}\\
    d\xi_t &= &\frac{1}{8F^2}\expect{v_{\xi_t}, d\ket{\psi_t}[\xi_t,\theta_t]} ,
\end{eqnarray}
where $d\ket{\psi_t}[\xi_t,\theta_t]$ is the evolution of $\ket{\xi_t,\theta_t}$ under the Stratonovich filter of Eq.\ (\ref{eqn:double_pass_SSE_stratonovich}).   After converting the resulting expressions back to an It\^{o} equation from its Stratonovich form, the resulting two-parameter filter on Gaussian states is given by \cite{Chase:2009b}
\begin{eqnarray}\label{eqn:projected_filter_ito}
		    d\theta_t &= &\left[\omega_t 
		               - \frac{M}{4}e^{-16F\xi_t}\sin(2\theta_t) 
		               + 2F\sqrt{KM}\sin{\theta_t}\right]dt  \\
		  && \quad\quad \nonumber
		                -\left[\sqrt{M}e^{-8F\xi_t}\cos{\theta_t} + \sqrt{K}\right]dW_t   \\
		    d\xi_t &= & \frac{M}{4}e^{-8F\xi_t}\cos^2{\theta_t}dt,  \nonumber
\end{eqnarray}
where the innovations are also expressed within the Gaussian state approximation of the filtered expectation value $\expect{\Fz}_t = - F \sin \theta_t$ to give
\begin{eqnarray}
	dW_t & = & dZ_t - 2\sqrt{M}\expect{\Fz}_tdt \nonumber \\
	& = & dZ_t + 2F\sqrt{M}\sin{\theta_t}dt .
\end{eqnarray}

\begin{figure}
\begin{center}
\includegraphics{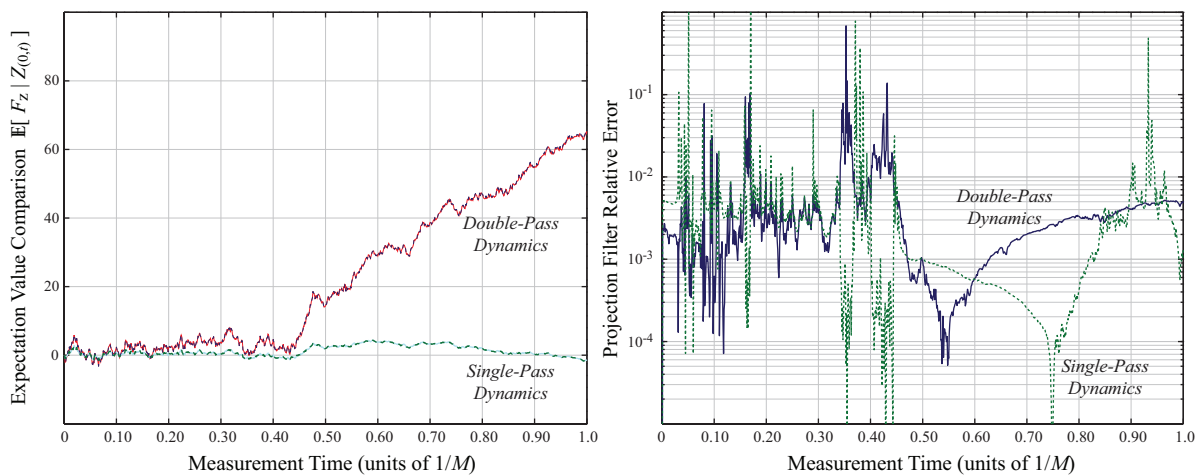}
\end{center}
\vspace{-4mm}
\caption{(color online) Comparison between the projection and full quantum filtering equations for the measurement trajectories in Fig.\ (\ref{figure:nonlinear_dynamics}). In plot (a), $\pi_t(\Fz) = \mathbbm{E}[ \Fz | Z_t]$ is computed for a simulated measurement trajectory using both the projection filter Eq.\ (\ref{eqn:projected_filter_ito}) and the full filter Eq.\ (\ref{eqn:double_pass_SSE_2}) under the same conditions as used in Fig.\ (\ref{figure:nonlinear_dynamics}).  The two filters given nearly indistinguishable resulsts.  Plot (b) shows the relative error $|(\pi_t^\mathrm{(exact)}(\Fz) - \pi_t^\mathrm{(proj)} ) /\pi_t^\mathrm{(exact)}(\Fz) | $ for both the double-pass and single-pass evolutions. \label{figure:filter_comparison}}
\end{figure}

Figure (\ref{figure:filter_comparison}) illustrates the accuracy of the projection filter by comparing the conditional expectation value $\pi_t^\mathrm{(proj)}(\Fz) = \mathbbm{E}_t[ \Fz | \hat{Z}_{(0,t)}]$ computing from Eq.\ (\ref{eqn:projected_filter_ito}) with the conditional expectation $\pi_t^\mathrm{(exact)}$ obtained from Eq.\ (\ref{eqn:double_pass_SSE_2}) under the same conditions used to generate Fig.\ (\ref{figure:nonlinear_dynamics}).  The relative error in the projection filter is found to be below 1\%, except for regions where the filter crosses zero (which accentuates the relative error).  One would expect the projection filter to perform increasingly better with larger values of $F$, but the comparison in the figure is limited to values for which Eq.\ (\ref{eqn:double_pass_SSE_2}) can be evaluated in practice.

%%%%%%%%%%%%%%%%

\section{Quantum Parameter Estimation and Atomic Magnetometry} \label{section:parameter_estimation}

Since the effective dynamics of the double-pass atomic system in Fig.\ (\ref{figure:schematic}) are seen to amplify rotations about the $y$-axis, it raises the question whether such a configuration provides an improved capability to detect external influences that drive $y$-axis rotations in the atoms.   In atomic magnetometry \cite{Budker:2002a,Romalis:2003a,Geremia:2003a}, for example, a $y$-axis magnetic field $\mathbf{B} = B\,  \vec{\mathbf{y}}$ is typically detected by preparing an atomic sample into the $x$-polarized initial state $\ket{F,+F}_\mathrm{x}$, allowing it to undergo Larmor precession until some time $t$ and then inferring the value of $B$ by measuring the $z$-component of the atomic spin to determine the amount of rotation.  For very precise measurements, uncertainty $\delta \tilde{B}$ in the estimated value $\tilde{B}$ of the field is dominated by quantum fluctuations in the observations performed on the atomic sample \cite{Helstrom:1976a,Braunstein:1994a,Wineland:1993a}.  

Magnetometry fits naturally into the coherent feedback plus continuous measurement structure developed in Sections (\ref{section:physical_system}) and (\ref{section:quantum_filter}).  The atomic sample couples to the magnetic field via the Zeeman Hamiltonian
\begin{equation} \label{eqn:larmor_hamiltonian}
	\hat{H} =  - \hbar \gamma B \Fy,
\end{equation}
where $\gamma$ is the atomic gyromagnetic ratio, which has precisely the structure considered in the derivation of the Gaussian state projection filter with $\omega_t = \gamma B$ in Eq.\ (\ref{eqn:y_hamiltonian}).  Measurements of the $z$-component of the spin reveals information about Larmor precession, which should then in-turn enable estimation of $B$ by devising a suitable estimation procedure [as addressed in Section (\ref{section:quantum_particle_filter})].

\subsection{Lower-Bounds in Quantum Parameter Estimation}  

The first problem at hand, however, is to ascertain whether the positive feedback amplification of Eq.\ (\ref{eqn:larmor_hamiltonian}) offers a potential advantage for determining the value of $B$ as compared to an estimation procedure that relies on only a single-pass measurement configuration \cite{Geremia:2003a}.  At first brush, it would seem that amplification of the Larmor dynamics such as observed in Fig.\ (\ref{figure:nonlinear_dynamics}) should make the influence of the field easier to observe, especially for small values $| B | \ll 1$.  Amplification of Eq.\ (\ref{eqn:larmor_hamiltonian}) is, however, not the entire story because an amplification of $\Fz$ occurs even in the absence of an applied field.  A magnetic field estimator must be able to distinguish the amplification of the magnetic signal from amplification of the coherent state spin projection noise \cite{Geremia:2003a,Wineland:1993a,Wineland:1994a}.

The potential for positive feedback amplification to improve the parameter estimation uncertainty can be analyzed via the quantum Cram\'{e}r-Rao inequality \cite{Helstrom:1976a,Holevo:1982a,Braunstein:1994a,Braunstein:1995a}, which places an information-theoretic lower bound on the units-corrected mean-square deviation of any estimator $\tilde{B}_t$ from the true value of the field $B$,
\begin{equation} \label{eqn:CRError}
    \delta \tilde{B}_t = \left\langle\left(\frac{\tilde{B}}
    {\abs{d\expect{\tilde{B}}/dB}} - B\right)^2 \right\rangle^{1/2}
\end{equation}
in terms of the dynamics
\begin{equation} \label{eqn:cramer_rao}
    \delta \tilde{B}_t \geq \mathrm{tr} \left[  4 \hat{\rho}_t(B) \left(\frac{\partial \hat{\rho}_t(B)}{\partial B}\right)^2
    	\right] ^{- \frac{1}{2}}.
\end{equation}
The behavior of the estimator uncertainty depends on the characteristics of the quantum states used to compute the expectation value in Eq.\ (\ref{eqn:cramer_rao}) and also the nature of the induced dynamics \cite{Boixo:2007a}.  If one considers only Hamiltonian evolution under Eq.\ (\ref{eqn:larmor_hamiltonian}) and does not permit quantum entanglement between the different atoms in the probe, the Cram\'{e}r-Rao inequality then yields the so-called shotnoise uncertainty \cite{Budker:2002a,Geremia:2003a}
\begin{equation} \label{eqn:deltaB_shotnoise}
	\delta \tilde{B}_t^\mathrm{(SN)} = \frac{1} { \gamma t \sqrt{2 F }} ,
\end{equation}
whose characteristic $1/\sqrt{N}$ scaling is a byproduct of the projection noise $\langle \Delta \Fz\rangle =\sqrt{F/2}$ for a spin coherent state \cite{Wineland:1993a,Wineland:1994a}.  If, however, entangled states are allowed to evolve under Eq.\ (\ref{eqn:larmor_hamiltonian}), then the Cram\'{e}r-Rao inequality yields the so-called Heisenberg uncertainty   \cite{Braunstein:1994a,Braunstein:1995a,Geremia:2003a,Leibfried:2004a}. 
\begin{equation} \label{eqn:deltaB_heisenberg}
	\delta \tilde{B}_t^\mathrm{(HL)} = \frac{1} { 2 \gamma t F} ,
\end{equation}
which offers a quadratic improvement over shotnoise as a function of the number of atoms.  Furthermore, if one goes beyond the structure of Eq.\ (\ref{eqn:larmor_hamiltonian}) by implementing dynamics that are non-linear in the collective spin operator, it is possible to surpass the $1/N$ scaling \cite{Boixo:2007a} as a consequence of multi-body effects \cite{Boixo:2007a,Rey:2007a}. 

\subsection{Numerical Evaluation of Cram\'{e}r-Rao Bound}
\label{section:cramer_rao_simulations}

A lower-bound on the magnetic field estimation uncertainty possible under the dynamics generated by the double-pass configuration in Fig.\ (\ref{figure:schematic}) can be obtained by integrating the quantum filter in Eq.\ (\ref{eqn:double_pass_SSE}) and computing a finite difference approximation to the operator derivative $\partial \hat{\rho}_t(B)/\partial B$ in Eq.\ (\ref{eqn:cramer_rao}).   Such a finite difference approximation can be constructed by co-evolving three trajectories, $\hat{\rho}_t(B)$, $\hat{\rho}_t(B+\delta )$, and $\hat{\rho}_t(B-\delta )$, and calculating
\begin{equation} 
    \left< \left(\frac{\partial \hat{\rho}_B(t_s)}{\partial B}\right)^2 \right>
        \approx \mathrm{tr} \left[ \hat{\rho}_t(B) \left(\frac{\hat{\rho}_t(B+\delta )  
                - \hat{\rho}_t(B-\delta ) }{2\delta }\right)^2 \right]
                \label{eqn:cr_fd}
\end{equation}   
for $\delta  \ll 1$.  The estimator uncertainty bound $\delta \tilde{B}_t$ obtained in this manner is conditioned on the random measurement realization driving the innovations process $dW_t$ for any single integration of the quantum filter.  Averaging Eq.\ (\ref{eqn:cr_fd}) over many such measurement realizations then provides the unconditional bound corresponding to the double-pass magnetic field estimator.

\begin{figure}
\begin{center}
\includegraphics{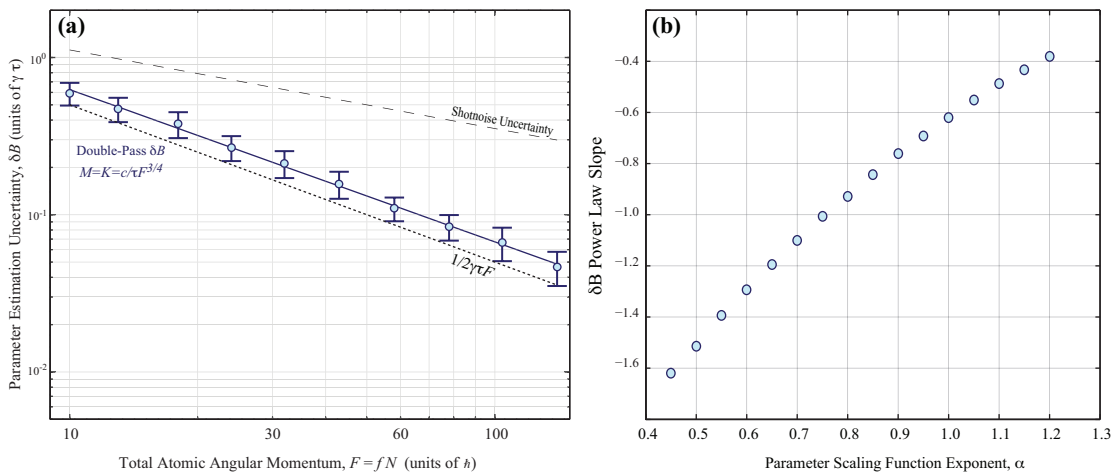}
\end{center}
\vspace{-4mm}
\caption{(color online) Plot of a lower bound on the estimator uncertainty for atomic magnetometry performed according to the measurement configuration in Fig.\ (\ref{figure:schematic}) as determined from a numerical evaluation of the quantum Cram\'{e}r-Rao inequality.  Plot (a) demonstrates that it should be possible to achieve $1/F$ scaling of the magnetometer sensitivity by adjusting the measurement parameters $M$ and $K$ as a function $F$ according to Eq.\ (\ref{eqn:MK_scaling}) with $\alpha =0.77$.  A fit to the data points in plot (a) gives a scaling law of $F^{-0.98}$, which is consistent with $1/F$ to within the accuracy of the fit.  Plot (b) illustrates that is possible to surpass the Heisenberg uncertainty scaling using different for different values of $\alpha$.  \label{figure:cramer_rao}}
\end{figure}

A thorough numerical investigation of the behavior of $\delta \tilde{B}_t$ computed from the Cram\'{e}r-Rao inequality Eq.\ (\ref{eqn:cramer_rao}) evaluated using Eq.\ (\ref{eqn:cr_fd}) reveals that the estimator performance depends very highly on the values of the coupling parameters $M$ and $K$, and that the best values of $M$ and $K$ depend on the spin size $F$ \cite{Chase:2009a,Chase:2009b}.  This is not surprising, as even in a classical amplifier, achieving increased signal to noise often requires selecting an optimal value for the amplifier gain.  The current situation is no different.  By conducting large sets of simulations, we identified empirically that scaling the values of $M$ and $K$ according to the functional form \cite{Chase:2009a}
\begin{equation} \label{eqn:MK_scaling}
	M = K = \frac{c}{ t_\mathrm{final} F^\alpha}, 
\end{equation}
where $c$ and $\alpha$ are constants, leads to a power-law scaling for the uncertainty bound $\delta \tilde{B}_{t_\mathrm{final}} \sim 1/N^k$.  As demonstrated by the data points in Fig.\ (\ref{figure:cramer_rao}a), it is possible to achieve $1/N$ scaling (to within a small prefactor offset) by setting $\alpha \approx 0.77$. A linear fit to the Cram\'{e}r-Rao data (on a log-log scale) gives $\delta\tilde{B}_{t_\mathrm{final}} \sim F^{-0.98}$, which reproduces the Heisenberg scaling to within the precision of the fit.  The data in Fig.\ (\ref{figure:cramer_rao}b) shows the slope of a linear fit of $\log_{10}\delta\tilde{B}_{t_\mathrm{final}}$ to $\log_{10}F$ (i.e., a slope of $k=-1$ corresponds to the Heisenberg uncertainty scaling) for different values of $\alpha$ in Eq.\ (\ref{eqn:MK_scaling}).    This plot suggests that it should be possible to surpass the Heisenberg ``limit'' as well via a double-pass measurement, however, these calculations provide no guarantee that the Gaussian state approximation continues to hold for larger values of $M$ and $K$.  In any given measurement configuration, it is essential to ascertain whether or not the Gaussian family provides a sufficient approximation to the actual atomic state.

Extrapolating the values of $M$ and $K$ that would be required to achieve a sensitivity $\delta\tilde{B}_{t_\mathrm{final}}$ corresponding to the Heisenberg uncertainty scaling Eq.\ (\ref{eqn:deltaB_heisenberg}) with $N \sim 10^8$ (which is a conservative value for typical experiments) implies that one would only require $M t_\mathrm{final} \sim 10^{-7}$, which suggests that $1/N$ scaling can be achieved by an extremely weak measurement, with no appreciable generation of conditional spin-squeezing \cite{Chase:2009a}.   This suggests that it may be much easier than previously believed to improve the sensitivity of an atomic magnetometer in practice--- scalings beyond $1/\sqrt{N}$ are made possible through the use of amplified dynamics instead of the preparation of an entangled state of the atoms.  In the case of small $M$ and $K$, the Gaussian family should provide a good representation of the atomic state.

\subsection{Quantum Particle Filter} \label{section:quantum_particle_filter}

The Cram\'{e}r-Rao inequality provides an ideal lower-bound on the sensitivity of a parameter estimator, but does not generally yield a constructive procedure for implementing parameter estimation.   It is, however, possible to obtain a magnetic field estimation procedure for the double-pass measurement using a technique referred to as \textit{quantum particle filtering} \cite{Chase:2009b}.  The concept of quantum particle filtering was developed in Ref.~\cite{Chase:2008b}.  It operates by embedding a classical probability distribution for the parameter to be estimated into a quantum probability space \cite{Chase:2008b,Chase:2009b}.  This is accomplished by promoting the classical parameter to be estimated--- in this case $B$--- to a diagonal operator
\begin{equation}
    B \mapsto \hat{B} = \int B \ketbra{B}{B} dB \in \mathcal{H}_B, 
\end{equation}
where $\mathcal{H}_B$ is an auxiliary Hilbert space for the parameter with basis states satisfying $\hat{B}\ket{B} = B\ket{B}$ and $\braket{B}{B'} = \delta(B-B')$.   All atomic operators and states, which are associated with the atomic Hilbert space $\mathcal{H}_A$, act as the identity on $\mathcal{H}_B$, e.g. $\Fz \mapsto I \otimes \Fz$.  The only operator which joins the two spaces is the parameter coupling Hamiltonian, \begin{equation}
    \hat{H} \mapsto -\hbar\gamma \hat{B}\otimes\Fy .
\end{equation}

The standard methods of quantum filtering theory are then employed to obtain the best least-squares estimate of the parameter via the conditional expectation
\begin{equation}
	\tilde{B}_t = \pi_t(\hat{B}) = \mathbbm{E}[ j_t (\hat{B}) | Z_{(0,t)} ]
\end{equation}
following the approach detailed in Section (\ref{section:quantum_filter}).  Since $\hat{B}$ corresponds to a classical parameter, the marginal density matrix $\mathrm{tr}_{\mathcal{H}_\mathrm{A}}[\hat{\rho}_t]$ should be diagonal in the basis of $\hat{B}$ so that it represents a classical probability distribution.  Therefore the conditional density operator propagated by the adjoint form of the quantum filtering equation
\begin{equation} \label{eqn:ensemble_continuous_form}
    \hat{\rho}_t = \int  p_t(B) \ketbra{B}{B} \otimes \hat{\rho}_t^{(B)} dB 
\end{equation}  
where $p_t(B) = P(B | Z_{(0,t)})$ is precisely the conditional probability density for $B$.  The density function describes a continuous variable, but in practice is discretized using a weighted set of point masses or \emph{particles}:
\begin{equation} \label{eqn:approximate_density}
    p_t(B) \approx \sum_{i=1}^N p_t^{(i)} \delta(B - B_i), 
\end{equation}
an approximation which can be made arbitrarily accurate in the limit of $N \rightarrow \infty$.  Under this discretization, the conditional density operator in Eq.\ (\ref{eqn:ensemble_continuous_form}) takes the form
\begin{equation} \label{eqn:discrete_ensemble_rho}
    \hat{\rho}_t^E = \sum_{i = 1}^N p_t^{(i)} \ketbra{B_i}{B_i} \otimes \hat{\rho}_t^{(B_i)}, 
\end{equation}
where each of the $N$ triples $\{p_t^{(i)}, B_i, \hat{\rho}_t^{(B_i)} \}$ is called a \emph{quantum particle}, giving rise to the name of the parameter estimation procedure.  The parameter estimate can then be obtained from the approximate density in Eq.\ (\ref{eqn:approximate_density}) either by taking the most probable $B$ value, i.e., the $B_i$ that maximizes $p_t^{(i)}$, but generally by calculating the expected value of $\hat{B}$
\begin{eqnarray}
    \tilde{B}_t & = & \pi_t(\hat{B}) \\
    & =& \mathbbm{E}[\hat{B}_t | Z_{(0,t)}] \approx \sum_{i=1}^{N} p_t^{(i)} B_i \nonumber
\end{eqnarray}
and the parameter estimation uncertainty is obtained from the variance of the conditional density
\begin{equation} \label{eqn:estimator_variance}
    \delta \tilde{B}^2_t =\expect{\hat{B}_t^2} - \tilde{B}_t^2 
    	= \sum_{i=1}^{N} p_t^{(i)} B_i^2 - \tilde{B}_t^2 .
\end{equation}

\subsubsection{Gaussian State Particle Filter.}

The quantum particle filter for the double-pass system within the Gaussian family restriction is found by substituting $\hat{\rho}_t$ into the extended double-pass filter and then projecting onto the manifold Eq.\ (\ref{eqn:gaussian_state_parameterization}).  The resulting particle filter is given by
\begin{eqnarray}
		    dp_t^{(i)} &= & 2 F \sqrt{M} \left[ \sin{\theta_t^{(B_i)} } 
		        - \sum_{j=1}^N p_t^{(j)} \sin{\theta_t^{(B_j)}} \right] p_t^{(i)}dW_t 
		          \label{eqn:projection_particle_filter} \\
		    d\theta_t^{(B_i)} &= &\left[ \gamma B_i 
		               - \frac{M}{4} \exp \left(-16F\xi_t^{(B_i)} \right) \sin 2\theta_t^{(B_i)}  
		               + 2F\sqrt{KM}\sin{\theta_t^{(B_i)}}\right]dt  \nonumber \\
		  && \quad\quad \nonumber
		                -\left[\sqrt{M}\exp\left(-8F\xi_t^{(B_i)} \right)\cos{\theta_t^{(B_i)}} 
		                	+ \sqrt{K}\right]dW_t   \\
		    d\xi_t^{(B_i)} &= & \frac{M}{4}\exp \left( -8F\xi_t^{(B_i)} \right)
		    	\cos^2{\theta_t^{(B_i)}}dt,  \nonumber
\end{eqnarray}
with the modified innovations process
\begin{equation}
		    dW_t = dZ_t  + 2 F\sqrt{M}\sum_{i=1}^Np_t^{(i)} \sin{\theta_t^{(B_i)}}dt.
\end{equation}
Evidently, each particle's quantum state $\hat{\rho}_t^{(B_i)}$ evolves under the projection filter corresponding to the known field $B = B_i$, while the innovations process $dW_t$ serves to couple the different particles.  The form of this filter, which involves several terms that are nonlinear in the variables $\theta_t$ and $\xi_t$ suggest that, even if the state is Gaussian and the estimator density $p_t(B)$ is of a Gaussian form, it is still unlikely that Kalman filtering methods \cite{Geremia:2003a} can be applied as a result of the nonlinear dynamics.

\begin{figure}
\begin{center}
\includegraphics{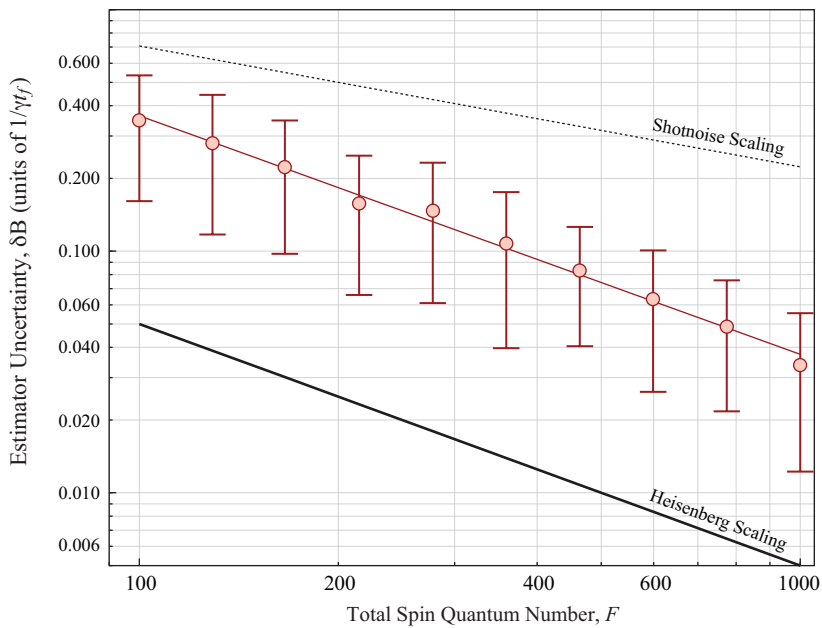}
\end{center}
\vspace{-4mm}
\caption{(color online) Numerical simulation of the Gaussian projection particle filter for estimating the value of an external magnetic field that uses coherent amplification of the magnetic dynamics to achieve a sensitivity that scales as $1/F$ without the need for spin squeezing.  \label{figure:estimator_scaling}}
\end{figure}

\subsubsection{Parameter Estimation Simulations.} We performed simulations of the Gaussian projection particle filter magnetic field estimator Eq.\ (\ref{eqn:projection_particle_filter}) to assess its performance relative to the Cram\'{e}r-Rao predictions in Fig.\ (\ref{figure:cramer_rao}).  As is often the case when analyzing detection limits, we took the true value of the field to be $B=0$, and evolved a statistical ensemble of measurement realizations under Eq.\ (\ref{eqn:projection_particle_filter}).  For our simulations, we utilized $N_\mathrm{p}=10^4$ particles and took the prior distribution over $B$ to be uniform 
\begin{equation}
	p_0^{(i)} = \frac{1}{N_\mathrm{p}}
\end{equation}
with the particle values $B_i$ placed on a uniform grid with domain $\pm \sqrt{D}$ centered around $B=0$.  That is, the particles have the values
\begin{equation}
	B = \left\{ - \sqrt{D}, - \sqrt{D} + dB, \ldots, \sqrt{D} -dB, \sqrt{D} \right\} 
\end{equation}
with $dB = 2 \sqrt{D} / N_\mathrm{p}$.  For our dynamics, we considered $\gamma=1$ with the initial quantum state given by $\ket{\theta_0,\xi_0}$ with $\theta_0 =0 $ and $\xi_0 = 0$, i.e., the $x$-polarized spin coherent state.  The coupling parameters, $M$ and $K$, were scaled according to Eq.\ (\ref{eqn:MK_scaling}) using $c = 0.5888$ and $\alpha = 0.77$, whose values were chosen such that the estimator uncertainty should scale as $1/F$ with the size of the spin system.

The results of the simulations are given in Figure (\ref{figure:estimator_scaling}), which plots the estimator uncertainty $\delta \tilde{B}$ obtained from Eq.\ (\ref{eqn:estimator_variance}) as a function of $F$.   The data points indicate the average value of $\delta \tilde{B}$ over the statistical ensemble of 10,000 measurement realizations, and the error bars indicate the standard deviation of $\delta \tilde{B}$ over that ensemble.  For reference, the shotnoise and Heisenberg uncertainty scalings are also plotted on the axes, making it apparent that the projection particle filter clearly outperforms shotnoise.  A numerical fit of the projection particle filter scaling gives $\delta \tilde{B} \sim F^{-0.98}$, which has the same power law slope as the Heisenberg scaling to within the precision of the fit.  The absolute magnitude of the parameter estimation uncertainty $\delta \tilde{B}$ exhibits, however, a displacement from that of the Heisenberg uncertainty limit.  This is not surprising, since the Cram\'{e}-Rao inequality provides a theoretical lower bound, and there is no guarantee that an estimator (even an optimal estimator) is capable of achieving that bound.    As expected, based on the analysis of Section (\ref{section:cramer_rao_simulations}), the atomic state not only remains Gaussian, but also undergoes extremely little squeezing.  For the measurement realizations considered here and scaling under Eq.\ (\ref{eqn:MK_scaling}), typical values of $\xi$ were found be $\xi \approx 1.2 \times 10^{-3}$ for $F=100$ and $\xi \approx 2.0 \times 10^{-4}$ for $F=1000$.  Since $\xi=0$ corresponds to a spin coherent state, this constitutes an improvement over shotnoise without appreciable generation of entanglement between atoms.   Extrapolating our results to experimentally viable values of $F \sim 10^8$, or more, suggests that coherent feedback amplification of the Larmor dynamics enables $1/F$ uncertainty scaling in parameter estimation without the need for spin squeezing, making the approach attractive for practical application.

\begin{figure}
\begin{center}
\includegraphics{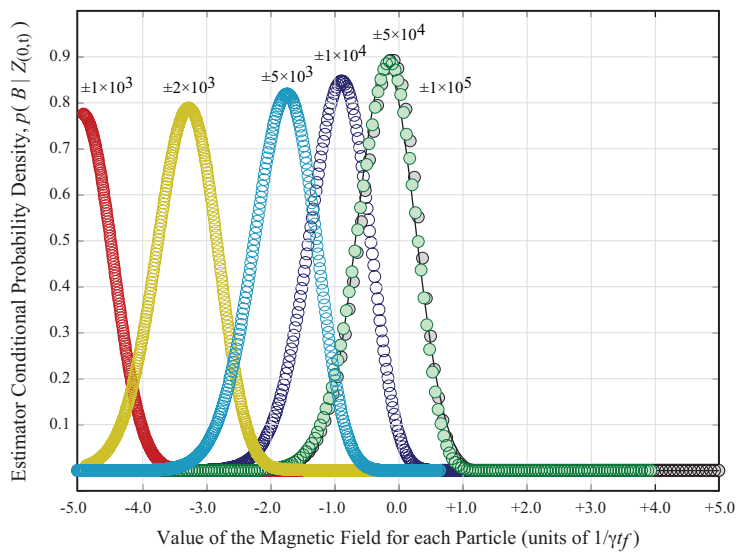}
\end{center}
\vspace{-4mm}
\caption{(color online) The conditional probability density produced by the projection particle filter exhibits a bias unless a sufficient number of particles is employed and the prior distribution is sufficiently wide.  The figure demonstrates reduction of bias and convergence of the estimator as the width of the prior distribution is increased.  In each case, the prior distribution is taken to be uniform over the range of $B$-values given by $\pm \sqrt{D}$.  The filter is seen to converge (for this measurement realization) as $D$ is increased from $1\times 10^3$ to $1\times 10^5$.  The solid line is a fit of the evolved conditional probability density to the converged normal distribution.  \label{figure:estimator_convergence}}
\end{figure}

Several important factors remain be to be solved, however, before the procedure developed here could be successfully incorporated into an actual laboratory setting.   As in any situation where a continuous probability distribution is discretized, as required for numerical analysis, there is the potential to introduce bias into the estimator outcome, even if the underlying analytic methods are bias-free \cite{Chase:2008b,Doucet:2001}.  During the stochastic evolution of the particle filter, especially the early transient period of the evolution, it is possible for the distribution mean to vary significantly.  Noise realizations that couple to particles which lie far away from the true value of $B$ can develop a bias if the tails of the distribution exceed the domain to any significant degree.  Such a bias might offset $\tilde{B}$ from the true value of $B$ by an amount that exceeds the estimator uncertainty $\delta \tilde{B}$, and thus prevent the estimator from achieving its expected uncertainty.   In an experiment, such a bias would be diagnosed by observing that the sample variance of the estimator $\mathbbm{E}[ (\tilde{B} - B)^2 ]$ exceeds that predicted by Eq.\ (\ref{eqn:estimator_variance}).  

We have observed such a bias in our simulations, and have found that care must be taken to minimize the introduction of bias by using a sufficiently wide prior distribution on $B$.   Figure (\ref{figure:estimator_convergence}) demonstrates the reduction of estimator bias as the width of the prior distribution is increased from $D = 1\times10^3$ to $D  = 1\times 10^5$.  With increasing $D$, the mean of the estimator distribution converges toward the Gaussian distribution illustrated by the solid line in Fig.\ (\ref{figure:estimator_convergence}).  We have also found that the value of $D$ required for such convergence exhibits some variability with respect to the particular measurement realization.  This is not unexpected, as the degree to which the filter must incorporate particles that lie near the domain boundaries depends upon the stochastic propagation, and different noise realizations cause the filter to explore $B$-space differently.    

Increasing the domain by taking $D \gg 1$ does not come without a cost. For a fixed number of particles, the filter grid size $dB = 2 \sqrt{D} / N_\mathrm{p}$ increases with $D$.  For a fixed value of $F$, a smaller fraction of the particles have non-negligable weight at the end of the filter evolution as $D$ is increased.  This reduced resolution is also evident in evident in Fig.\ (\ref{figure:estimator_convergence})-- the larger domain sizes coincide with reduced resolution of $p(B|Z_{(0,t)})$.  The number of particles must be chosen such that there is adequate resolution $p(B|Z_{(0,t)})$ to determine its mean and variance with confidence.  As the number of atoms increases, $F$ the width of $p(B|Z_{(0,t)})$ decreases, suggesting that the number of particles must also be increased to maintain adequate resolution.  As $N_\mathrm{p}$ increases, it becomes increasingly time-consuming to integrate the filtering equations, and this could prove limiting in an actual magnetometry application.  Fortunately, there are a number of potential methods one might use to improve the efficiency of the filter by reducing the effective number of particles.  For example, particles whose weight drops below a given threshold could be removed from the integration, or a more sophisticated basis than the delta function representation Eq.\ (\ref{eqn:approximate_density}) could be used to resolve $p(B|Z_{(0,t)})$.  While a full exploration of the numerical methods suitable for optimizing the performance is beyond the scope of this work, such practical considerations are likely to play an essential role in deploying particle filters in a laboratory experiment.

\section{Conclusions} \label{section:conclusion}

We have presented a method for quantum parameter estimation, especially suited toward atomic magnetometry, based on the amplification of Hamiltonian dynamics via coherent positive feedback.   We also explored the behavior of our feedback system using information-theoretic parameter estimation bounds, via the Cram\'{e}r-Rao inequality.  To render our approach more suitable for incorporation into an actual laboratory setting, we employed methods from quantum filtering theory and differential geometry to obtain a simplified estimator based on a two-parameter family of Gaussian atomic spin states.  Finally, we demonstrated that it should be possible for an atomic magnetometer to achieve a sensitivity scaling comparable to the Heisenberg scaling without the need for entanglement.   Our results suggest that effective nonlinear dynamics, such as those obtained here through positive coherent feedback, are likely to play as important a role in the field of quantum parameter estimation as classical amplifiers play in classical metrology.

\ack
We thank Rob Cook and Ben Baragiola for many discussions. This work was supported by the National Science Foundation (PHY-0639994).  Please visit http://qmc.phys.unm.edu for additional information, including Matlab/C++ code and data files used to generate the figures presented in this work.

\section*{References}

\bibliography{NJPArticle}
\end{document}